\documentclass[12pt,preprint]{aastex}

\renewcommand{\min}{\mbox{$^m$}}

\def\Sec{${}^{\prime\prime}$\llap{.}}
\def\min{${}^{\prime}$}
\def\Min{${}^{\prime}$\llap{.}}
\def\ltsim{ \,{}^<_\sim\, }

\newcommand{\strom}{\mbox{Str\"omgren~}}
\newcommand{\omc}{\mbox{$\omega$ Cen~}}

\shorttitle{Reddening distribution in {\boldmath $\omega$}\,Cen} 
\shortauthors{Calamida et al.}

\begin{document}



\title{Reddening distribution across the center of the globular cluster $\omega$ Centauri\altaffilmark{1}}

\author{
A. Calamida\altaffilmark{2,3},
P. B. Stetson\altaffilmark{4,16,17},
G. Bono\altaffilmark{3},
L. M. Freyhammer\altaffilmark{5,6},
F. Grundahl\altaffilmark{7},
M. Hilker\altaffilmark{8},
M. I. Andersen\altaffilmark{9} 
R. Buonanno\altaffilmark{2,3},
S. Cassisi\altaffilmark{10}, 
C. E. Corsi\altaffilmark{3},
M. Dall'Ora \altaffilmark{3,11},
M. Del Principe\altaffilmark{10},
I. Ferraro\altaffilmark{3}, 
M. Monelli\altaffilmark{3,12},
A. Munteanu\altaffilmark{13},
M. Nonino\altaffilmark{14},
A. M. Piersimoni\altaffilmark{10},
A. Pietrinferni\altaffilmark{10},
L. Pulone\altaffilmark{3},
T. Richtler\altaffilmark{15} 
}

\altaffiltext{1}{
Based in part on observations collected with Danish and NTT telescopes 
operated at ESO La Silla}

\altaffiltext{2}{
Universita' di Roma Tor Vergata, Via della Ricerca Scientifica 1,
00133 Rome, Italy}

\altaffiltext{3}{
INAF-Osservatorio Astronomico di Roma, Via Frascati 33, 00040, Monte Porzio 
Catone, Italy; bono, buonanno, calamida, corsi, pulone@mporzio.astro.it}

\altaffiltext{4}{
Dominion Astrophysical Observatory, Herzberg Institute of Astrophysics,
National Research Council, 5071 West Saanich Road, Victoria, BC V9E~2E7,
Canada; Peter.Stetson@nrc-cnrc.gc.ca}

\altaffiltext{5}{
Royal Observatory of Belgium, Ringlaan 3, B-1180 Brussels, Belgium;
lfreyham@vub.ac.be}

\altaffiltext{6}{
Vrije Universiteit Brussel, OBSS/WE, Pleinlaan 2, B-1050 Brussels, Belgium;
lfreyham@vub.ac.be}

\altaffiltext{7}{Dept. of Physics \& Astronomy, Aarhus Univ.,
Ny Munkegade, 8000 Aarhus C, Denmark; fgj@phys.au.dk
}

\altaffiltext{8}{Sternwarte Bonn, Auf dem H\"ugel 71, D-53121 
Bonn, Germany; mhilker@astro.uni-bonn.de
}

\altaffiltext{9}{Astrophysikalisches Institut Potsdam, Sternwarte 16, 
D-14482 Potsdam, Germany; mandersen@aip.de 
}

\altaffiltext{10}{INAF-Osservatorio Astronomico di Collurania, via M. Maggini, 
64100 Teramo, Italy; cassisi@te.astro.it, milena@te.astro.it, 
piersimoni@te.astro.it, adriano@te.astro.it 
}

\altaffiltext{11}{INAF - Osservatorio Astronomico di Capodimonte,
via Moiariello 16, 80131 Napoli; dallora@na.astro.it
}

\altaffiltext{12}{Instituto de Astrofisica de Canarias, Via Lactea,
E38200 La Laguna, Tenerife, Spain; monelli@iac.es
}

\altaffiltext{13}{Universitat Politecnica de Catalunya, Spain;
andreea.munteanu@upf.edu
}

\altaffiltext{14}{INAF-Osservatorio Astronomico di Trieste, via G.B. Tiepolo 11,
40131 Trieste, Italy; nonino@ts.astro.it  
} 

\altaffiltext{15}{Universidad de Concepcion, Departamento de Fisica, Casilla 
106-C, Concepcion, Chile; tom@coma.cfm.udec.cl  
}

\altaffiltext{16}{Guest User, Canadian Astronomy Data Centre, which is operated
by the Herzberg Institute of Astrophysics, National Research Council of Canada.}

\altaffiltext{17}{Guest Investigator of the UK Astronomy Data Centre.}

\date{\centering drafted \today\ / Received / Accepted }

\begin{abstract}

We present new medium-band {\it uvby\/} \strom and broad-band {\it VI\/}
photometry for the central regions of the globular cluster $\omega$ Cen.  From
this photometry we have obtained differential reddening estimates relative to
two other globular clusters (M$\,$13 and NGC$\,$288) using a
metallicity-independent, reddening-free temperature index,
$[c]\equiv(u-v)-(v-b) - 0.2(b-y)$, for hot horizontal-branch (HB) stars ($T_e\ge
8,500\,$K). We estimate color excesses of these hot HB stars using optical and
near-infrared colors, and find clumpy extinction variations of almost a factor
of two within the area of the cluster core.  In particular, the greatest density
of more highly reddened objects appears to be shifted along the right ascension
axis when compared with less reddened ones. These findings complicate 
photometric efforts to investigate the star formation history of $\omega$ Cen.     

\end{abstract}



\section{Introduction}\label{introduction}

The difficulty of obtaining absolute reddening estimates toward globular
clusters (GCs) has plagued distance determinations, and in turn absolute
age estimates \citep{ren91, cast99, gra03}. Uncertain reddening also
affects the comparison between theory and observations \citep{zoc00}.
This problem becomes even more severe for GCs affected by differential reddening 
\citep{pier02, ste03}.  Several methods have been suggested to provide
robust reddening estimates based on either medium- and broad-band photometry \\
\citep{we85, ste91, twa93, kov03} or high-resolution spectra \cite{car04}.  
However, current methods to estimate cluster reddenings may be
affected by systematic uncertainties \citep{pier02, zoc01}. This 
applies not only to GCs with very low reddening such as M3 \citep{cac05}, but
also to moderately reddened clusters like $\omega$ Cen. Current reddening
estimates toward $\omega$ Cen cluster around $E(B-V)=0.11\pm0.02$ 
\citep{lub02, thom01}. 
However, the reddening map of Schlegel et al.\ (1998) indicates
reddening variations of $\sim0.02$ across the body of the cluster, while 2MASS
data \citep{law03} show a very clumpy reddening distribution outside 1${}^\circ$
(100~pc) from the cluster center.  

In a recent investigation \citep{fre05}, we presented accurate optical
({\it BRI\/}) and near-infrared ({\it JK\/}) photometry for stars in $\omega$
Cen. Adopting a metal-intermediate chemical composition ($0.003<Z<0.015$) and
an age coeval with the bulk of the $\omega$ Cen stars, we found a
plausible fit of the anomalous red giant branch ($\omega$3, Lee et al. 1999) 
if we increase the distance modulus by $\Delta \mu=0.2$, and the reddening by
$\sim$ 0.03. This suggests that the $\omega$3 branch could
be a clump of stars located  $\sim 500$ pc beyond the main body of $\omega$ Cen,
and that there may be a clumpy reddening distribution across the cluster.
To further constrain the possibility of differential reddening
toward $\omega$ Cen, we decided to use hot HB stars ($T_e \ge 8,500$ K), since 
their relation between effective temperature and photometric color is minimally
affected by chemical composition (Webbink 1985).  We adopt a reddening-free
temperature index based on \strom photometry, $[c] = c_1-0.2(b-y)$, where $c_1=
(u-v)-(v-b)$ and $E(b-y)= 0.74 E(B-V)$ \citep{craw76,card89}, to estimate the
reddening of individual hot HB stars. The main advantage of this approach is
that the $[c]$ index is a very robust temperature indicator for these stars.  

\section{Observations and reddening estimates}
\label{observations}

A set of 110 $uvby$ \strom images of \omc were collected by L.~M.~Freyhammer in
April 1999 with the Danish Telescope (ESO, La Silla). The average seeing of
these images is $\sim$1\Sec5 and the field of view is 14\min$\times$14\min\ on
the cluster center.  These have been supplemented with 30 {\it uvby\/} images of
the Southwest quadrant of \omc collected by F.~Grundahl in April/June 1999 with
the same telescope, together with a sample of HD standard stars.  We also have
210 {\it vby\/} images collected with the Danish Telescope by Hilker and
collaborators in two observing runs (1993 and 1995, see Hilker 1999; Hilker \&
Richtler 2000).  The photometry was performed using DAOPHOTII/ALLSTAR/ALLFRAME
\citep{ste87, ste94}. 
The final merged star catalog includes $\approx 2\times 10^5$ stars. 
The typical photometric precision for faint hot HB stars is better 
than $\sim 0\fm03$ at $y \approx 19.5$ and better than $\sim 0\fm02$ at 
$u \approx 19$ mag.  The absolute calibration of 
Grundahl's data was based on $\sim$ 120 HD standard stars observed
over six non-consecutive nights. We defined local cluster standards 
to calibrate the remaining overlapping fields. A comparison 
between our calibrations and those
of Richter et al. (1999) indicates agreement in {\it vby\/} 
better than 0.02--0.03 mag. 
A detailed discussion of the photometry and calibration 
will be given in a future paper (Calamida et al. in preparation). 

Near-infrared $JK_s$ images of $\omega$ Cen collected in 2001/2002 
with SOFI@NTT (ESO, La Silla) were analyzed together with data from 
2000, available in the ESO archive. The merged
catalog contained $\sim$1$\times$10$^5$ stars. We also collected 
$VI$ images in 1999, with FORS1@VLT (standard-resolution 
mode, ESO, Paranal). The seeing was better than 1\Sec0
and the final catalog includes $>5\times10^5$ stars. 
The accuracy of the absolute zero-points is of the order of 
0.02-0.03 mag (Freyhammer et al. 2005).
To obtain differential reddening estimates for \omc we selected two GCs, 
specifically M$\,$13 (= NGC$\,$6205 = C1639+365, [Fe/H]=--1.54) and NGC$\,$288
(= C0050-268, [Fe/H]=--1.24), each marginally affected by reddening 
(E(B-V)=0.02, M$\,$13; E(B-V)=0.03, NGC$\,$288; Harris 2003), 
and each possessing an extended blue HB for which both {\it uvby\/} 
\strom $\,$ \citep{gru99} and {\it VI\/} photometry
exist.  Our {\it VI\/} photometry consists of both original and
archival data analyzed by PBS (Stetson 2000).  
Fig.~1 compares the $u-y$ vs $[c]$ (top)
and $V-I$ vs $[c]$ (bottom) relations of hot HB stars in \omc (dots),
M$\,$13 (diamonds), and NGC$\,$288 (triangles). This figure
shows that hot HB stars in \omc are systematically redder and display, at fixed
$[c]$, larger color dispersions ($\sigma_{u-y}=0.085$) compared to 
M$\,$13 ($\sigma_{u-y}=0.037$) and 
NGC$\,$288 ($\sigma_{u-y}=0.016$)\footnote{Omega Cen data plotted in 
Fig. 1 extend to [c] values higher than for the other two clusters. 
This does not appear to be the result of an overall shift of the 
cluster loci---the mid-lines appear to agree very well, suggesting 
that the absolute calibrations are consistent. Obviously, reddening 
variations might scatter stars up and down but not horizontally. 
HB stars in Omega Cen outnumber the other cluster samples by almost a factor
of ten. Therefore the random photometric error distribution is populated farther
out into the wings, and the apparent spread for $[c]\approx1.2$ of Omega Cen
stars ought be due to sample size.}. The spread in color might be due to 
variations in the reddening toward $\omega$~Cen, since the $[c]$ index is 
reddening-free and the colors of hot HB stars should 
marginally depend on metal abundance.    

To confirm that colors of hot HB stars are metal-independent 
we transformed theoretical Zero Age Horizontal Branches (ZAHBs) 
for three different
chemical compositions into the ($u-y$ vs $[c]$) observational plane using
the color-temperature transformations and bolometric corrections of
Castelli (2005, private communications). For more details concerning 
the input physics and the
adopted evolutionary parameters see Pietrinferni et al.\ (2004). 
Fig.~2 shows that HB structures hotter than 8,500 K ($u-y < 1.7$) are 
scarcely affected by metal abundance; the same applies to the 
other \strom and broad-band colors. 
Note that current scenario depends at most very weakly on an alpha-element
enhancement, since according to current evolutionary predictions only stellar
structures cooler than 6,500 K appear to be affected (see Fig. 1 in Cassisi
et al. 2004). 

Supported by this evidence, we estimated the differential reddening for each hot
HB star in \omc by fitting fiducial sequences to the hot HB stars in M$\,$13 and
in NGC$\,$288 in the $u-y$, $v-y$, and $b-y$ vs $[c]$ planes and in the ($V-I$)
vs $[c]$ plane. From the extinction model of Cardelli et al. (1989), we derived
extinction coefficients for the \strom bands (column 4 of Table 1), and the mean
differential reddening in the four colors (columns 1 and 2 of Table 1).  Allowing
for the estimated reddening of NGC$\,$288 and M$\,$13, we find
a mean reddening for \omc $\sim\langle E(B-V)\rangle= 0.13\pm 0.04$ (from the
NGC$\,$288 comparison) or $\sim \langle E(B-V)\rangle= 0.10\pm 0.03$ (from M$\,$13).
These estimates agree 
well with values available in the literature. The typical photometric precision 
in the colors is $\le 0.03$ and $\le 0.02$ for hot HB stars 
in M~13 and in NGC~288, respectively. Note that the absolute reddening 
of $\omega$~Cen calibrated against M$\,$13 is smaller than that
calibrated against NGC$\,$288, suggesting that the published 
reddening for M$\,$13 might be slightly low (i.e., it should 
be $\sim 0.04$ instead of 0.02) or the reddening of NGC$\,$288 might be 
high (i.e., it should be  $\sim 0.01$ instead of 0.03).
However, the concordance between the color excesses 
estimated from the different colors
suggests that the difference does {\it not\/} stem primarily from 
photometric calibration errors.  

Although the new mean reddening value for $\omega$~Cen agrees well 
with literature values, reddening estimates for individual hot 
HB stars present a star-to-star scatter larger than our error budget. 
In particular, the reddening seen in the $u-y$ color ranges from
$E(u-y)\sim 0.05$ to $E(u-y)\sim 0.28$ ($0.03 \ltsim E(B-V)\ltsim
0.15$), while the reddening from the $V-I$ color ranges from
$E(V-I)\sim0.08$ to $E(V-I)\sim0.17$ ($0.06 \ltsim E(B-V)\ltsim
0.13$). Similar values obtain for the $v-y$ and $b-y$ colors.
This suggests variable reddening toward the cluster core with a
dispersion $\sigma_{E(B-V)} \approx 0.03$ and supports the results
of Cannon \& Stobie (1973), who found reddening
variations in \omc\ with $\sigma_{E(B-V)} \approx 0.03-0.05$. 
Evolutionary effects in the (color) vs $[c]$ planes could
only produce bluer hot HB stars at a given fixed
luminosity.  However, these colors are not very sensitive to gravity at
these temperatures, so luminosity evolution should be {\it along\/} the
(color)-$[c]$ relationships, and not {\it away\/} from them.
To make this key point more 
clear, the inset in Fig. 2 shows that the off-ZAHB evolution, core-He 
burning phase, of selected hot HB models (Y=0.246, Z=0.001, M=0.508, 
0.53, 0.58 $M_\odot$) takes place along the (color)-[c] relation.
This means that only a reddening variation or a non-stellar spectral
energy distribution can move an object off the (color)-[c] relation.      

To further constrain this hypothesis, we have also investigated
the color distribution of hot HB stars in the $u-J$ and $u-K$ colors.
Again, we estimated the ridge lines of hot HB stars in the different (color)
vs $[c]$ planes and the color-distances of
individual objects from the ridge lines. Fig.~3 shows the distribution for all
six colors together with Gaussian fits (solid lines). 
To assess whether individual color excesses might be due to photometric errors, 
we investigated the star-by-star correlations among the different colors.
Fig.~4 shows that the color excesses are well correlated
in both medium and broad-band photometry. Note that the outliers
in $u-J$ and $u-K$ are stars close to the cluster center (d $\le$ 1\Min5) whose
photometry could be affected by crowding.

Finally, we investigated the spatial distribution of less-, more-, and 
average-reddened HB stars (respectively, outside and within
$\pm1\sigma$).  Fig.~5 shows that the distribution of these 
objects is clumpy.  
In particular, at the largest radii shown, less-reddened HB stars 
(blue dots) outnumber the more-reddened ones (red dots), especially in the
Northwest, Northeast, and Southeast quadrants. Moreover, less-reddened
stars on the East (left, $X > 0$) side of the cluster outnumber those
on the West side, and a less pronounced although still real absence of
blue dots is seen in the Northwest (top right) quadrant and at small 
radii. There is also evidence for a particular shortage of more-reddened
stars with $X > 250$ and an unusually low dispersion in reddening in
the Southwest quadrant, with few blue OR red dots present at
larger radii.

\section{Discussion and final remarks} \label{Discussion}

We have derived homogeneous and accurate {\it
uvby\/} \strom and {\it VI\/} photometry for 
\omc.  An empirical method based on hot HB stars provides robust
differential reddening estimates relative to two globulars marginally affected
by reddening, NGC$\,$288 and M$\,$13. In all colors, the reddening estimates for
individual stars show variations of at least a factor of two within our field.
This supports the suggestion of differential
reddening toward \omc by Dickens \& Caldwell (1988) and 
by Freyhammer et al. (2005) from comparisons
between evolutionary predictions and accurate multi-band photometry, and
also the mild extinction variations detected 
by Minniti et al. (1992) from linear polarization.
Unfortunately, the latter study does not cover the cluster core. Finally, 
circumstellar emission has been recently detected in five \omc red giants 
near the tip of the RGB and AGB by Boyer et al. (2004) 
in Spitzer Space Telescope IRAC images ranging from 3.6 to 8.0 microns. 

The overall distributions on the sky of more- and less-reddened stars 
are rather different, suggesting that the foreground material
is clumped on arcminute angular scales. 
However, more- and less-reddened stars are also found
close together on the sky, suggesting that either:
(a)~the absorbing material also displays structural features on arcsecond scales, or
(b)~some of the absorbing material is within the cluster, and we are seeing
front-to-back differences as well as foreground gradients.
Our results show that \omc is indeed subject to variable reddening,
which should be taken into account in photometric investigations. 
Omega Cen is different from M22, where the {\it entire\/} color dispersion
among RGB stars may be due to patchy foreground extinction (Richter et
al. 1999). However, dust in front of or inside \omc does complicate the efforts to
uncover the details of its formation history and metallicity spread, and more
work to quantify the extinction pattern is highly desirable.
Thus, these findings should be independently confirmed with different stellar
tracers. For instance, high-resolution spectra across $H_\alpha$ or perhaps 
the Na-D lines plus accurate multiband photometry of RGB stars 
would be very useful to disentangle reddening 
and metallicity effects.


\acknowledgements
We wish to tank an anonymous referees for his/her suggestions that
helped us to improve the content and the readability of the 
manuscript. Part of this research was supported by  
``IAP P5/36'' Interuniversity Attraction Poles Programme of 
the Belgian FOSTCA, the FWO, and the Flemish Ministry for
Foreign Policy, European Affairs, Science and Technology, 
(BIL 01/2), and by COFIN 2003 from MIUR.



\clearpage

\begin{figure}[!h]
\begin{center}
\label{fig1}
\includegraphics[height=0.37\textheight,width=0.45\textwidth]{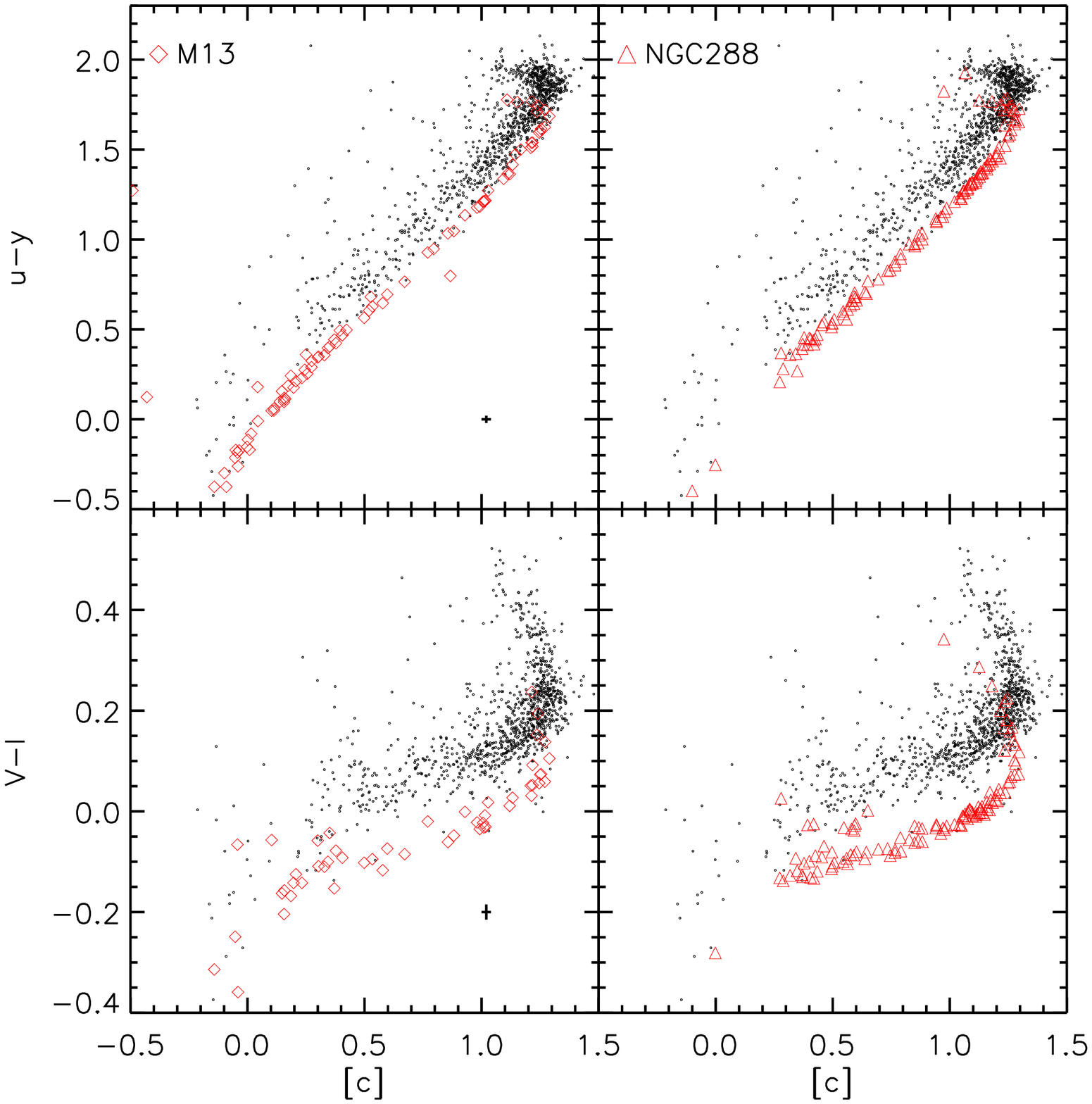}
\vspace*{0.2truecm}
\caption{Top - Comparison in the $u-y$ vs $[c]$ plane between hot 
HB stars in \omc (dots, $\sim$1000 stars), in M13 (diamonds, $\sim$80 stars), 
and in NGC288 (triangles, $\sim$120 stars). Error bars show intrinsic 
photometric and calibration errors. Bottom - Same as the top, but in the 
$V-I$ vs $[c]$ plane.}
\end{center}
\end{figure}

\clearpage

\begin{figure}[!h]
\begin{center}
\label{fig2}
\includegraphics[height=0.25\textheight,width=0.45\textwidth]{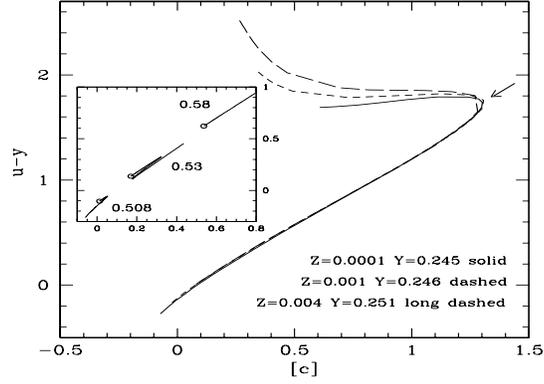}
\vspace*{0.2truecm}
\caption{Predicted ZAHBs for different chemical compositions. 
The helium (Y) and the metal abundance (Z) are labeled. The arrow 
points HB structures with $T_e\approx 8,500 K$. The inset shows 
the off-ZAHB evolution, core-He burning, of three selected HB 
models for Z=0.001.}
\end{center}
\end{figure}

\clearpage

\begin{figure}[!h]
\begin{center}
\label{fig3}
\includegraphics[height=0.35\textheight,width=0.45\textwidth]{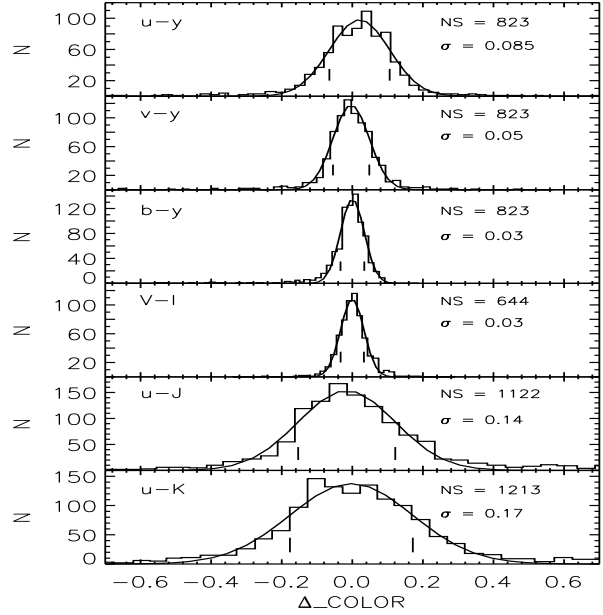}
\vspace*{0.2truecm}
\caption{Color excess for individual hot HB stars in \omc 
relative to the empirical ridge lines in different (color) vs $[c]$ planes.
The \strom sample was selected by requiring a measurement in all 
four bands. The number of hot HB stars and the $\sigma$ of the 
Gaussian fit are labeled. Vertical bars mark $\pm 1 \sigma$.} 
\end{center}
\end{figure}

\clearpage

\begin{figure}[!h]
\begin{center}
\label{fig4}
\includegraphics[height=0.275\textheight,width=0.425\textwidth]{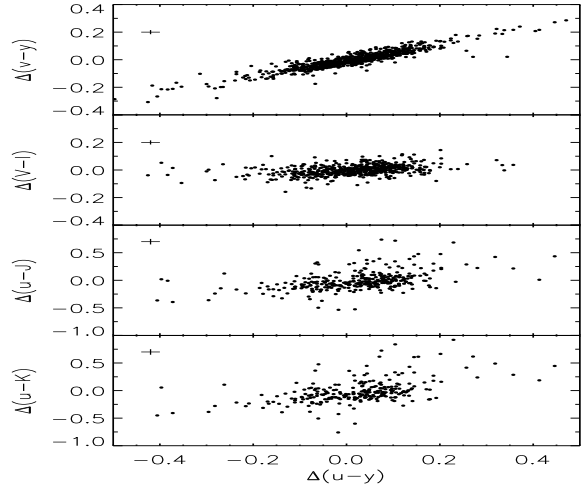}
\vspace*{0.2truecm}
\caption{Correlation between the color excess of hot HB stars based
on both optical and NIR photometry. Error bars show intrinsic photometric 
and calibration errors.
}
\end{center}
\end{figure}

\clearpage

\begin{figure}[!h]
\begin{center}
\label{fig5}
\includegraphics[height=0.40\textheight,width=0.45\textwidth]{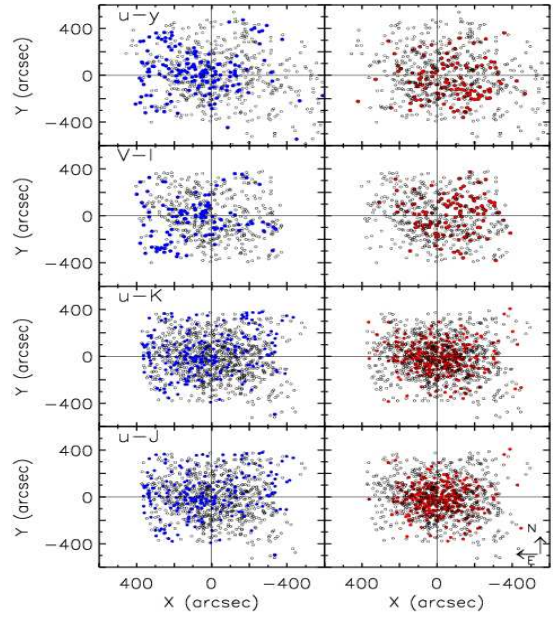}
\vspace*{0.30truecm}
\caption{Spatial distributions of hot HB stars in \omc. Objects that 
are 1 $\sigma$ redder/bluer than the ridge line are plotted as red 
and blue dots, while the normal-reddened (within 1 $\sigma$) HB stars 
as open circles.}
\end{center}
\end{figure}

\clearpage

\begin{deluxetable}{lcrccc} 
\tabletypesize{\scriptsize}
\tablecaption{Differential reddening estimates for hot HB stars 
in \omc \label{tab1}} 
\tablewidth{0pt}
\tablehead{
\colhead{Mean Red./[NGC288]\tablenotemark{a}}  &
\colhead{Mean Red./[M13]\tablenotemark{b}}  &
\colhead{NS\tablenotemark{c}}  &
\colhead{Ext. Coef.\tablenotemark{d}}  &
\colhead{$\langle E(B-V)\rangle$/[NGC288]\tablenotemark{e}} &
\colhead{$\langle E(B-V)\rangle$/[M13]\tablenotemark{f}} 
} 
\startdata 
$\langle E(u-y)\rangle$=0.18 (0.16/0.10) & 0.15 (0.19/0.10) & 1018 & 1.84 & 0.10$\pm$0.03 & 0.08$\pm$0.03 \\
$\langle E(v-y)\rangle$=0.12 (0.08/0.05) & 0.10 (0.08/0.05) &  932 & 1.33 & 0.09$\pm$0.02 & 0.08$\pm$0.02 \\
$\langle E(b-y)\rangle$=0.08 (0.07/0.03) & 0.05 (0.07/0.03) & 1021 & 0.70 & 0.11$\pm$0.02 & 0.08$\pm$0.02 \\
$\langle E(V-I)\rangle$=0.13 (0.06/0.04) & 0.12 (0.06/0.04) &  970 & 1.30 & 0.10$\pm$0.02 & 0.09$\pm$0.03 \\
\enddata  
\tablenotetext{a,b}{Mean differential reddening relative to NGC288 and to M13. 
The number in parentheses gives the standard deviation of the reddening 
distribution and the $\sigma$ of the Gaussian fit.}
\tablenotetext{c}{Number of hot HB stars in \omc adopted to estimate the 
differential reddening.}
\tablenotetext{d}{Extinction coefficients, E(CI)/E($B-V$), according to 
Cardelli et al. (1989).}
\tablenotetext{e,f}{Mean differential reddening, 
$\langle \Delta E(B-V)\rangle$, relative to NGC288 and to M13.}
\end{deluxetable}

\end{document}